# A new approach towards the self-adaptability of Service-Oriented Architectures to the context based on workflow

Faîçal Felhi

Department of Computer Sciences
High Institute of Management, BESTMOD Laboratory
Tunis, Tunisia

Jalel Akaichi

Department of Computer Sciences
High Institute of Management, BESTMOD Laboratory
Tunis, Tunisia

*Abstract*— **Distributed information systems are needed to be autonomous, heterogeneous and adaptable to the context. This is the reason why they resort Web services based on SOA Based on the advanced technology of SOA. These technologies can evolve in a dynamic environment in a well-defined context and according to events automatically, such as time, temperature, location, authentication, etc... This is what we call self-adaptability to context. In this paper, we are interested in improving the different needs of this criterion and we propose a new approach towards a self-adaptability of SOA to the context based on workflow and showing the feasibility of this approach by integration the workflow under a platform and test this integration by a case study.**

## I. INTRODUCTION

System Information (SI) must meet some specific constraints surrounding context adaptation in the case of ubiquitous computing [22]. This is particularly the case in areas of industry relating to many domains such as RFID (Radio Frequency Identification) [19][42], e-Learning ect... For this case, considerable approaches related to adaptability with different modes of implementation such as: AOP (Aspect Oriented Programming)[10]. This aspect used by various platforms on the goal to adapt the Web service (WS) [30] to the context dynamic changes of environment. Web services, like any other middleware technologies, aim to provide mechanisms to bridge heterogeneous platforms, allowing data to flow across various programs. The WS technology looks very similar to what most middleware technologies looks like. The emergence of Web services as a model for integrating heterogeneous Web information has opened up new possibilities of interaction and adaptability to context when offered more potential for interoperability. However, from a set of requirements on SOA (Service Oriented Architecture in English) [6], and to provide self adaptation to the context of Web services, we need to integrate more generic connector that takes into account all ambient or distant events.

Based on technology platforms to adapt of Web services to the context, and also gain benefit from the advantages of Web services. These platforms deal with "simple" adaptation to functional and technical exchange purpose it is by no means clustering adaptation to ambient context year. These systems must be used in different contexts depending on the environment of the user profile and the terminal to use. One of the major problems of such systems relates to the context adaptation. It is important that applications adapt to their surroundings [23]. The adaptation software can take many forms, and we refer to system adaptable when a user can interact with the system and, through them, modify and customize. An adaptive system identifies a situation and adapts. Activation of this system changes can be caused by human intervention or a number of observations.

The main goal of context execution is to allow the application to manage the external situations that affect its quality of service seen by the user. Thus, the application must be adaptable to disappearances and appearances of devices on the network, for example, or all sorts of technical failures, malicious or unusual expense. Our goal is to study the adaptability with a presentation of platforms to adapt their operating principles, and trying to improve the SOA to empower the Web services to be self adaptive to the contexts. In our paper we propose an approach for self adaptation of SOA to the context.

The rest of this paper is organized as follows: In Section 2, we review previous research on adaptability of Web services as well as other related issues. Section 3 proposes the needs of self adaptability of SOA to the context. In section 4 we present our approach for self adaptability of SOA to the context and illustrated our work by feasibility and a case study. Finally, we summarize our work and discuss future research in Section 5.

## II. RELATED WORKS

Based on aspects, AspectWerkz [12], supporting weaving at loading, offers various means of expression of cut points. It offers different methods to specialize or customize the middleware according to the code on the works. Furthermore, it supports different various methods of expression (XML and Java). DynamicTAO [8] have a particularity is offer via a file different aspects of the ORB [2] in the form of strategy design pattern. The system configuration is offer via a file (specifies the different strategies applied by the ORB before launching the system). Based on our work, we conclude that this platform suffers like all conventional middleware for the inability to reconfigure the ORB during execution time. DynamicTAO, address aspects such as: competition, safety and supervision, and can provide a set of interfaces allowing users to reconfigure its structure.

Each WS has an Interface Definition Language, namely WSDL (Web Service Description Language) [31], that is responsible for the message payload, itself described with the equally famous protocol SOAP (Object Access Protocol) [32], while data structures are explained by XML (eXtended Markup Language) [33]. Very often, WS are stored in UDDI (Universal Description Discovery and Integration) [34] registry. Based on aspects and Web services, Charfi and al. approach [1] propose a framework that provides support for middleware BPEL (Business Process Execution Language) [37] engines. The authors apply the concepts of deployment descriptor and container for the Web service composition. Ferraz Tomaz and al. approach [24] proposed a tool for weaving aspects for a simple adaptability of the Web services, implementing aspects of the services as loosely coupled, where aspects are woven dynamically. In this approach, aspects are themselves Web services, thus they are independent of languages and platforms. This approach provides a mechanism, based on the AOP for Web services, to dynamically adapt to different policies of use. This approach has two major limitations related to its architecture and its implementation. These limitations are: the dependence of the runtime architecture of Web services and dependence on aspects of language.

Mehdi Ben Hmida approach [18] extended the solution proposed by [24] to specify BPEL processes adaptable, that is to say, the adaptability of complex services. This specification allows the generation of customers that adapt to dynamic changes on the server side. Changes made to the service can lead to further exchanges of information between client and server that were not initially planned during implementation of the client application. These new interactions normally cause runtime errors of the customer. This approach is based on process algebras to dynamically generate customers. Process algebras manage the interactions between a BPEL process and its customers, this by specifying formally the interaction protocol (abstract BPEL) and automatically generating a client who is successfully communicating with the service. This approach overcomes the limitations in the dynamic modification of a process that can lead to a change in the pattern of interaction with the client and will fit the client and server parts. Hence the need to extend the semantic aspects and Web services, which resulted in the ASW (Aspect Service Weaver). Aspects are themselves loosely coupled Web services, they are independent of languages and platforms, but, this approach has limitations. Adaptation to context is not taken into account, that is to say, if an event occurred during a search on a Web service, this approach does not take into account this event.

In the other approaches we find those based on context adaptation. The ambient computing encourages the proliferation of associated devices. A key aspect of the ambient computing is its invisibility. Users perceive the features but do not see the devices that provide these features. Adaptability and evolution of software in these devices becomes an asset to their condition of use. From studies by [4] we can summarize the descriptions of the following platforms. Aura [5] is a context-sensitive middleware that enables the design of mobile applications. Aura's goal is to provide each user with a set of implicit processing services and information. ExORB [21] demonstrates the ability of a middleware to support its configurability, the possibility of putting it-to-day improvement. ExORB examine in particular middleware for mobile phones. Such devices require a middleware on which it is possible to configure new software, to improve the software already built without manual intervention by the end user. ExORB purposed of contributing to the construction of middleware services for strengthening the main features of configurability, the ability to update and improvement. DoAmI (Domain-Specific Ambient Intelligence) [17] deals with the dynamic aggregation of distributed services. Its offers an architectural model that relies on a service-oriented middleware for integration and activation of services at runtime.

CORTEX (Co-operating Real-time sentient objects: architecture and Experimental Evaluation) [9] aims to build a middleware component-based applications to accommodate influenced by the external environment, particularly in the transportation field. CORTEX needs and capabilities of local services for local decision making. The system participates in a cooperative global system. Thus, it has a system for collecting real-time environment. In addition, local systems can be equipped with additional features such as consideration of traffic lights or a mechanism to allow pedestrians (presence sensor, obstacle avoidance). CORTEX defines objects aware. They are moving objects that behave independently and are responsible for interactions with the physical environment. These behaviors are based on sensor inputs and the internal state of each object. To address the problems of coordination, control, adaptability and scalability, CORTEX provides a first programming paradigm using objects aware. The discovery mechanism is not well detailed by the authors and does not measure the scalability of the architecture. In addition, it lacks the tools to do a self adaptation.

WComp [14][15][16] represent the implementation of experimental models presented in the work of RAINBOW (research team of the I3S laboratory, hosted by University of Nice - Sophia Antipolis). This is a platform for lightweight components for service composition SLCA (Service Lightweight Component Architecture) which enables the design of ambient computing applications by assembling software components, orchestrating access to services through infrastructure devices from ambient. WComp supports protocols such as UPnP (Universal Plug and Play) [38] and Web services, allowing components through the proxy to interact with them. WComp is a prototyping "development" environment for context-aware applications. The WComp Architecture is organized around Containers and Designers paradigms. The purpose of the Containers is to take into account system services required by Components of an assembly during runtime: instantiation, destruction of software Components and Connectors. The purpose of the Designers allows configuring assemblies of through Containers. To promote adaptation to context WComp uses Aspect [30] Assembly paradigm. Aspect Assemblies can either be selected by a user or fired by a context adaptation process. It uses a weaver that allows adding and or suppressing components. A container includes a set of (Beans) components and each bean has properties, input methods that use received input information, and output Methods to send to another bean, for instance, output information. Aspect Assemblies allow defining links between Beans by using input and output information. WComp uses UPnP (Plug and Play) technology to detect locally whether the device is active or not and to define input methods and sent events for each component. With this architecture WComp allows: i) managing devices heterogeneity and dynamic discovering by using UPnP, ii) events driven interactions with devices, iii) managing dynamic devices connection and disconnection (dynamic re configuration on run time) in infrastructure.

## III. ADAPTABILITY OF SOA TO THE CONTEXT AND NEEDS OF SELF ADAPTABILITY

### A. Adaptability of SOA to the context

The SOA offer great flexibility that is a great ability to functional and technical changes. Moreover, this type of architecture is most often used as Web services support, which provide the flexibility and interoperability expected, that is the ability to communicate between heterogeneous systems. When the SOA is based on Web services, is referred to as WSOA for (Web Services Oriented Architecture). The application in such information systems that incorporate SOA need to communicate across the exchange software (middleware or platforms). These middleware are the source of our work. It is on them that will think the same expectations in terms of flexibility, interoperability and adaptability.

The adaptability of a system [13] is the software mechanisms that achieve system changes. It is these mechanisms that modify the behavior of the system while preserving the properties of the system. The self-adaptability means the power to dynamically modify the behavior of a system in response to internal and external events. If the user has the possibility to adapt the interaction of these preferences, the interface is said to be configurable and adaptable. If the system is able to adapt his behavior to the needs (capabilities and preferences of the current user) during the interaction, with its capacities of perception and interpretation of the interaction and its context, the interface is called adaptive. The adaptivity of a system is how the system adapts. It consists of strategies to trigger changes in the system based on incentives.

The main goal of context [36] execution is to allow the application to manage the external situations that affect its quality of service seen by the user. Therefore, the application must be adaptable to disappearances and appearances of devices on the network, for example, or all sorts of technical failures, malicious or unusual expense. The context is not directly involved in the application, but it sets the execution environment of the application that is a subset of the software infrastructure. It is in fact available resources of the system at a given [26].

### B. Needs of self adaptability

The adaptation of service-oriented architectures is at the heart of building new applications.

A self-adaptable architecture is that it takes into account the business logic and HMI (Human Machine Interface). A workflow engine allows you to run a process defined elsewhere in the tool design process that accompanies it. This execution is sequential and is completely unconditional, conditional and is based on a set of rules defining the conditions of connection.

The workflow tools allow the definition of rules more or less sophisticated but still generally quite simple and few: Boolean operators, data fields of the process values entered by operators, properties of any documents involved in the process, etc…

The workflow engine can then connect to the rules engine to "know" what option to take during the course of a process. The rules engine also allows users to expose simple interfaces for generating these rules.

Following the research and the state of the art described above, lead us to determine the list of needs:

• An SOA for its flexibility, interoperability with Web services.

• Use of aspects for adaptation to changes that do not provide Web services.

• The use of such powerful concepts: reflexivity and / or MOP (Meta Object Protocol) [25] to gain flexibility and performance, or dynamic reconfiguration.

• Adaptation to the prevailing situation with mechanisms and platforms to capture the events and process them in real time, or in a near real time, by invoking the right service.

• Consideration of transverse functions such as security, log management, etc...

• Consideration of management rules intrinsically.

• The management of processes and data.

• Interoperable intelligent connectors that can be applied to the platform regardless of its technology.

## IV. SELF-ADAPTABILITY OF SOA TO THE CONTEXT BASED ON WORKFLOW

In this section we present the functional and technical architecture of our new approach for the self adaptability to the context of SOA based on the needs that we have already cited.

### A. Functional architecture WSOA with WWF

Fig. 1 shows our needs in terms of self adaptability of service oriented architecture to the context. We present in the following the functionality provided by our system:

• Decentralized and reconfiguration: our architecture is based on objects or components to make the dynamic reconfiguration of components using more advanced mechanisms. It qualifies the distribution of applications across multiple servers and not the increase in service levels. is a distributed architecture whose purpose is to deliver services to their audience and they will be accessible from any types of clients.

• Event and communication management: The events sent by the external environment are adapted to the context. This architecture must include an inference engine, which specifies the behavior of applications in a given context and uses the execution model events-conditions-actions. Management communication, use the mechanism based on events that are created dynamically during system operation. The analysis of the communication is based on modeling the communication events in composite services for ambient spaces.

• Weaving and generation: Weaving is the process that takes as input a set of aspects and a base application, and outputs an application whose behavior and structure are extended by aspects. Generating code corresponding to the component assembly, manipulation of the graphical representation of the application, generation of the executable code corresponding to the application. The weaving can Occur at compile time (Modifying the compiler), load time (Modifying the class loader) or runtime (Modifying the interpreter).

• Work flow and rules management: The workflow engine can then connect to the rules engine to "know" what option to take during the course of a process. The rules engine also allows users to expose simple interfaces for generating these rules.

• Composition and flow orchestration: enables the design of ambient computing applications by assembling software components and orchestration of access to services by devices from the ambient infrastructure.

• Security and administration: Offered by this system in treating the business logic from the workflow and rules.

• Discovery: Contextual resource discovery is the use of context data to discover other resources within the same context.

• Invocation ambient and distant services: The invocation of distant and ambient services is also permitted by this architecture using technologies dedicated to each type of invocation.

### B. Technical architecture WSOA with WWF

This architecture (Fig. 2) allows the structuring of technical capabilities and infrastructure in our new approach to SOA.

In this architecture, we present the different functionality by connectors well integrated within our system. What is specific in our approach, we propose to integrate connectors rules Engine that communicates with a workflow engine in our framework. In this rules engine we need to define the rules that manage the data flow to finally produce events providing services to the customer. The data management shall be provided from a component that specifies the service to send it to another component by assembling them that allows the management process, event management and orchestration services. Web services are invoked by remote proxy with their specific WSDL URL, as well as ambient services using specific technologies such as UPnP.

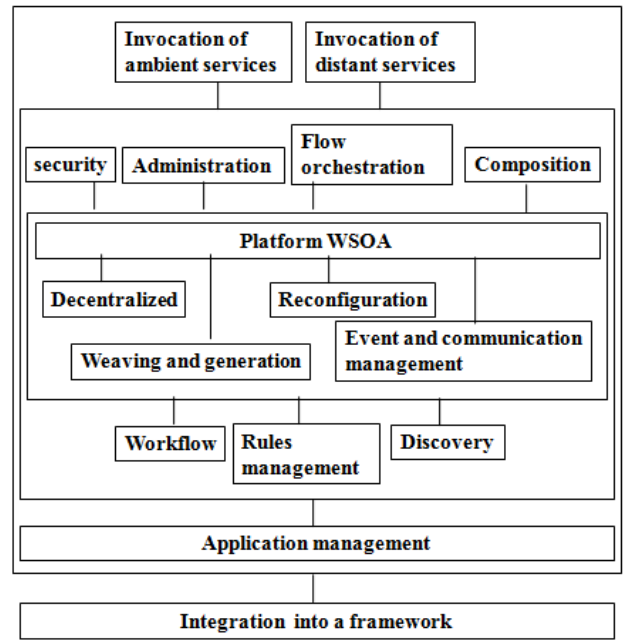

Figure 1.   Functional Architecture.

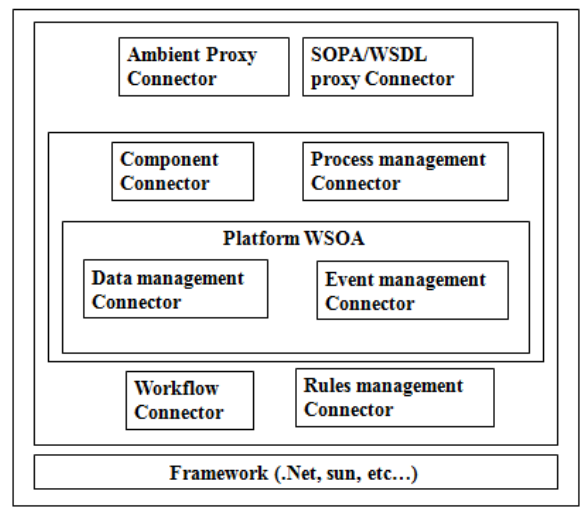

Figure 2.   Technical Architecture.

### C. Feasibility

In our research [26] [27] [28] we studied some platforms and in particular WComp and CORTEX, whose our main goal is to propose a self-adaptive SOA to the context. These studies have led to several implementations and case studies in different fields, such as, e-learning, smart house, RFID, etc... Result of these studies, we presented in our work [7] a proposal to an self-adaptable SOA when we've built a rules engine within WComp which can offer management rules that deal with business logic. Business logic can help in the development and optimization of these assemblies separating the events produced by the components defined in a WComp application.

Under WComp we have integrated a rule engine that can provide management rules that deal with business logic.

The rules engine can communicate with a workflow engine, which helps optimize and evolution of these assemblies separating the events produced by the components defined in an application WComp.WWF (Windows Workflow Foundation) [40][41] is a framework that allows users to create flow systems or user applications written for Windows Vista, Windows XP and Windows Server 2003 family. WWF can solve simple scenarios such as displaying user interface controls based on user input or complex scenarios encountered by large companies, such as order processing and inventory control. WWF treat the flux activation in business applications, the flow of pages the user interface, the flow of paper, user flows, mixed flows for applications based on services, and flow-driven rules business.

Our solution represented in Fig. 3 is based on the WWF under .Net that can solve simple scenarios such as displaying user interface controls based on user input or complex scenarios encountered by large enterprises. This integration solves simple scenarios such as displaying user interface controls based on user input or complex scenarios encountered by large enterprises.

In this architecture, except for the different needs initially used by WComp (service invocation ambient and remote data orchestration ...) describe above in the related word section, we integrated connectors rules engine that communicates with a workflow engine in framework .Net. In this rules engine we need to define the rules that manage the data flow to finally produce events providing services to the customer. The information shall be provided from a component that specifies the service to send it to another component by assembling them in a container through the language of Aspect of Assembly.

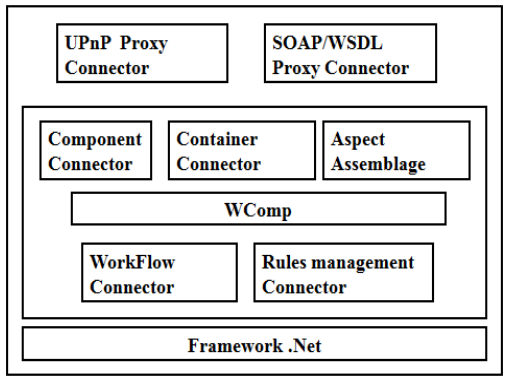

Figure 3.    Technical Architecture WComp integrating workflow.

### D.    Case Study

#### 1)    Description

We chose to take a sample case study of authentication. This authentication is supposed to capture with a RFID sensor using UPnP technology in WComp. This authentication can, thereafter, to monitor such access to a well determined. In our case, and to simulate this case study, we created a man-machine interface that captures a user authentication; this authentication is then verified based on business rules defined in XML, to finally show a message validation or invalidation of the value set as authentication.

#### 2)    Implementation

In a first step, we built the container WComp under sharpdevelop1.0.2a [39]. This container contains all components necessary to make the bean test authentication with an assembly between them. In a second step, we imported the container WComp sharpdevelop3.2 under a project to integrate the flows and rules. We chose to show the code (Fig. 4) of the rule that deals with the flow and data entered by the user. This code is in XML form.

In this block we defined as an example the rule that gives the exact value of authenticating a user. As shown in line 19, the exact value is "Felhi". This value is normally detected by RFID and simulated in this example by entering a "TexField", and for example displayed on a screen that is simulated by a "label".

```
15  <ns0:CodeBinaryOperatorExpression.Right>
16  <ns0:CodePrimitiveExpression>
17  <ns0:CodePrimitiveExpression.Value>
18  <ns1:String xmlns:ns1="clr-namespace:System;Assembly=mscorlib, Version=2.0.0.0, Culture=neutral,
    PublicKeyToken=b77a5c561934e089">Felhi</ns1:String>
19
20  </ns0:CodePrimitiveExpression.Value>
21  </ns0:CodePrimitiveExpression>
22  </ns0:CodeBinaryOperatorExpression.Right>
23  </ns0:CodeBinaryOperatorExpression>
24  </RuleExpressionCondition.Expression>
25  </RuleExpressionCondition>
26  </RuleDefinitions.Conditions>
27  </RuleDefinitions>
```

Figure 4.    Setting value.

## V.    CONCLUSIONS

In this paper we presented our new approach for a self-adaptability of service-oriented architectures to the context based on workflow by presenting the functional and technical architecture. In our solution we have given different features in terms of the needs of self-adaptability offered by the integration of workflow, which allows the management rules and a kind of security and administration of Web services. This solution which can offer management rules that deal with business logic. Business logic can help in the development and optimization of these assemblies separating the events produced by the components of Web services.

We have shown the feasibility of this approach by making use WComp platform and we integrated WWF workflow and we tested this integration through a case study.

This solution is a first step towards our problem of defining a service-oriented middleware self-adaptable to the context. We want in our future work added other connectors generic for the more needs of self adaptability.

### REFERENCES

[1]  A. Charfi and M. Mezini. "Aspect-Oriented Web Service Composition with AO4BPEL", (2004), 2nd European Conference on Web Services (ECOWS) Publisher, vol. 3250 of LNCS, Springer, pp. 168-182.

[2]  C. Schmidt Douglas and C. Cleeland. "Applying Patterns to Develop Extensible ORB MIddelware". (1999), IEEE Communications Publisher, Magazine Special Issue on Design Patterns Publisher.

[3]  C. Ullrich; K. Borau; H. Luo; X. Tan; L. Shen and R. Shen. Why Web 2.0 is Good for Learning and for Research: Principles and Prototypes. Proceedings of the 17th International World Wide Web Conference, ACM, 2008.

[4]   D. Cheung-Foo-Wo, M. Riveill, and J-Y Tigli, "Adaptation dynamique par tissage d'aspects d'assemblage", (2009), Thesis I3S, NICE-SOPHIA ANTIPOLIS.

[5]   D. Garlan, D. P. Siewiorek, A. Smailagic, and P. Steenkiste, "Aura : Toward distraction free pervasive computing", (2002), IEEE Pervasive Computing Publisher.

[6]   F. Curbera, R. Khalaf, N. Mukhi, "Quality of Service in SOA Environments". An Overview and Research Agenda (Quality of Service in SOA-Umgebungen). it - Information Technology 50(2): 99-107, 2008.

[7]   F. Felhi and J. Akaichi, "Adaptation of Web services to the context based on workflow: Approach for self-adaptation of service-oriented architectures to the context", (2012) International Journal of Web & Semantic Technology (IJWesT) Vol.3, No.4, October 2012 Publisher.

[8]   F. Kon, M. Roman, P. Liu, J. Mao, T. Yamane, L. Magalhaes, and R. Campbell, "Monitoring, security and dynamic configuration with the dynamictao reflective orb", (2000), Middleware'2000 Publisher, New York, USA.

[9]   G. Biegel and V. Cahill, "A framework for developing mobile, context-aware applications", (2004), 2nd IEEE Conference on Pervasive Computing and Communication, pp.361–365.

[10]  G. Kiczales, J. Lamping, C. Maeda, and C. Lopes, "Aspect-oriented programming", (1997), Proceedings European Conference on Object-Oriented Programming (ECOOP'97) , volume 1241, pp 220–242. Springer- Verlag, Berlin, Heidelberg, and New York.

[11]  G. Reese, "Cloud Application Architectures: Building Applications and Infrastructure in the Cloud", (2009), O'Reilly Publisher.

[12]  J. Bonér , "AspectWerkz: Dynamic AOP for Java", (2004) AOSD Publisher.

[13]  J-Y Tigli, S. Lavirotte, D. Cheung-Foo-Wo., (2003) "Mobilité et Enseignement à Distance" (special issue: TICE Seminar) International Journal of Information Science and Communication (ISDM) Publisher, number 10, ISSN 1265-499X.

[14]  J.-Y. Tigli, S. Lavirotte, G. Rey, V. Hourdin, and M. Riveill, "Lightweight Service Oriented Architecture for Pervasive Computing", (2009), IJCSI International Journal of Computer Science Issues, Vol. 4, No. 1, , ISSN (Online): 1694-0784, ISSN (Print): 1694-0814.

[15]  J.-Y. Tigli, S. Lavirotte, G. Rey, V. Hourdin, and M. Riveill, "Context-aware Authorization in Highly Dynamic Environments", (2009), IJCSI International Journal of Computer Science Issues, Vol. 4, No. 1, , ISSN (Online): 1694-0784, ISSN (Print): 1694-0814.

[16]  J.-Y. Tigli, S. Lavirotte, G. Rey, V. Hourdin, D. Cheung-Foo-Wo, E. Callegari, and M. Riveill. "WComp Middleware for Ubiquitous Computing: Aspects and Composite Event-based Web Services", (2009), Annals of Telecommunications, volume 64, n° 3-4, pp 197. ISSN 0003-4347.

[17]  M. Anastasopoulos, H. Klus, J. Koch, D. Niebuhr, and E. Werkman, "DoAmI – a middleware platform facilitating re-configuration in ubiquitous systems", (2006), System Support for Ubiquitous Computing Workshop. At the 8th Annual Conference on Ubiquitous Computing (Ubicomp) Publisher.

[18]  M. Ben Hmida, R. F. Tomaz, and V. Monfort, "Applying AOP concepts to increase Web services flexibility", (2006), Journal of Digital Information Management (JDIM) Publisher.

[19]  M. Hughes, "RFID tags for ambient intelligence: present solutions and future challenges", Joint sOc-EUSAI conference, Grenoble, 12-14 Octobre 2005.

[20]  M. Miller , "Cloud Computing: Web-Based Applications That Change the Way You Work and Collaborate Online", (2008), QUE Publisher.

[21]  M. Roman and N. Islam, "Dynamically programmable and reconfigurable middleware services", (2004), Middleware Publisher, Vol. 3231 in LNCS, pp. 372–396, Springer.

[22]  M. Weiser, "Some Computer Science Issues in Ubiquitous Computing", (1993), Communications of the ACM, vol. 36, no. 7, pp. 75–84.

[23]  N. Ferry and S. Lavirotte, (2008) "Adaptation Dynamique d'Applications au contexte", I3S laboratory, hosted by University of Nice - Sophia Antipolis.

[24]  R. Ferraz Tomaz, M. M. Ben Hmida, and V. Monfort, "Concrete Solutions for Web Services Adaptability Using Policies and Aspects", (2006), JDIM - Journal of Digital Information Management Publisher.

[25]  S. Chiba and T. Masuda, "Designing an Extensible Distributed Language with a Meta-Level Architecture", (1993), ECOOP'93, pp.482-501, LNCS 707, Springer-Verlag, Kaiserslautern, Germany.

[26]  S. Lavirotte, D. Lingrand, and J-Y Tigli. , "Définition du contexte : fonctions de coût et méthodes de sélection", (2005), Proceedings of the 2nd French-speaking conference on Mobility and ubiquity computing, pp. 9-12.

[27]  V. Monfort and F. Felhi, "Context Aware Management Platform to Invoke remote or local e Learning Services Application to Navigation and Fishing Simulator", International Symposium on Ambient Intelligence, (2010), ISAMI'10 Publisher, special volume in Advances in Intelligent and Soft Computing (Springer), Guimarães, Portugal.

[28]  V. Monfort, and F. Felhi, « A contextual approach to invoke intelligent house Services: an application to help physically handicapped persons » 1rst International Workshop on Recent Trends in SOA Based Information Systems in conjonction with ICEIS 2010, Funchal Madeira, Portugal, Juin 2010.

[29]  [29] V. Monfort, M. Khemaja, N. Ammari, and F. Felhi, « Using SaaS and Cloud computing For "On Demand" E Learning Services : Application to Navigation and Fishing Simulator », short paper in The 10th IEEE International Conference on Advanced Learning Technologies, July 5-7, 2010 Sousse, Tunisia.

[30]  Web URL - http://www.w3.org/TR/ws-arch/ (2004).

[31]  Web URL - http://www.w3.org/TR/wsdl20/ (2007).

[32]  Web URL - http://www.w3.org/TR/SOAP (2007).

[33]  Web URL - http://www.w3.org/XML/ (2012).

[34]  Web URL - http://www.uddi.org/pubs/uddi_v3.htm (2004).

[35]  Web URL - http://javaboutique.internet.com/articles/WSApplications/ (2012).

[36]  Web URL - http://www.larousse.fr/dictionnaires/francais, (2010).

[37]  Web URL - http://www6.software.ibm.com/software/developer/library/ws-bpel.pdf, (2003).

[38]  Web URL - http://www.upnp.org/, (2012).

[39]  Web URL - http://community.sharpdevelop.net/, ( 2011).

[40]  Web URL - http://www.bpmbulletin.com/2006/06/21/difference-entre-workflow-et-moteur-de-regle/, 2006.

[41]  Web URL -http://www.vdocsoftware.com/vdoc/easysite/InVDOC2010/news/innovation/agilite-regles-metiers, 2010.

[42]  Web URL - http://www.rfidfr.org/, 2012.

AUTHORS PROFILE

Faïçal Felhi received the master degree in Intelligent Information System in 2010. He is currently a PhD student in the BESTMOD laboratory in the high institute of management of Tunis. - Tunisia. He is actually teaching in the high institute of Computer and Mathematic of Monastir - Tunisia.

Jalel Akaichi received his PhD in Computer Science from the University of Sciences and Technologies of Lille (France) and then his Habilitation degree from the University of Tunis (Tunisia) where he is currently an Associate Professor in the Computer Science Department. He has published in international journals and conferences, and has served on the program committees of several international conferences and journals. He is currently the Chair of the Master Science in Business Intelligence. He visited and taught in many institutions such as the State University of New York, Worcester Polytechnic Institute, INSA-Lyon, University of Blaise Pascal, University of Lille 1, etc…